# Jet-Induced Nucleosynthesis in Misaligned Microquasars

Yousaf M. Butt[1], Thomas J. Maccarone[2], Nikos Prantzos[3]

[1]*Harvard-Smithsonian Center for Astrophysics, 60 Garden St., Cambridge, MA, USA*

[2]*Scuola Internazionale Superiore di Studi Avanzati, via Beirut 4, Trieste, ITALY*

[3] *Institut d'Astrophysique, 98bis Bd Arago, Paris, FRANCE*

**The jet axes and the orbital planes of microquasar systems are usually assumed to be approximately perpendicular, eventhough this is not currently an observational requirement. On the contrary, in one of the few systems where the relative orientations are well-constrained, V4641Sgr, the jet axis is known to lie not more than ~36$^o$ from the binary plane. Such a jet, lying close to the binary plane, and traveling at a significant fraction of the speed of light may periodically impact the secondary star initiating nuclear reactions on its surface. The integrated yield of such nuclear reactions over the age of the binary system (less the radiative mass loss) will detectably alter the elemental abundances of the companion star. This scenario may explain the anomalously high Li enhancements (roughly ~20-200 times the sun's photospheric value) seen in the companions of some black-hole X-ray binary systems. (Such enhancements are puzzling since Li nuclei are exceedingly fragile - being easily destroyed in the interiors of stars - and Li would be expected to be depleted rather than enhanced there.) Gamma-ray line signatures of the proposed process could include the 2.22 MeV neutron capture line as well as the 0.478 MeV $^7$Li$^*$ de-excitation line, both of which may be discernable with the *INTEGRAL* satellite if produced in an optically thin region during a large outburst. For very energetic jets, a relatively narrow neutral pion gamma-decay signature at 67.5 MeV could also be measurable with the *GLAST* satellite. We argue that about 10-20% of all microquasar systems ought to be sufficiently misaligned as to be undergoing the proposed jet-secondary impacts.**

Subject headings: black hole physics-- nuclear reactions, nucleosynthesis, abundances -- X-rays: binaries (V4641Sgr, GRO J1655-40) -- ISM: jets and outflows -- novae, cataclysmic variables

# I. INTRODUCTION

Microquasars are accreting galactic binary systems composed of a regular star orbiting a compact object - a black hole (BH) or neutron star (NS) - and are distinguished by having persistent or episodic jets of material which are ejected from close to the compact object's surface (event horizon) in opposite directions at relativistic speeds (Mirabel & Rodriguez 1999; Hjellming & Rupen 1995). It is believed that most accreting BH and NS X-ray binary (XB) systems are likely to exhibit microquasar behavior at some time during their life even if they have not yet been observed in this state.

Although it is not observationally required, the jet axis and the orbital plane of a given microquasar system are usually assumed to be approximately perpendicular. We argue that, to the contrary, a significant fraction of microquasar jets could lie close enough to the orbital plane so as to periodically impact their companion stars and initiate spallative and fusion nuclear reactions thereon. In this way the light-elements Li,Be and B (LiBeB) may be produced *in situ* on the secondary star. Galaxy-wide, the mass loss from the disks and secondary stars of all such binary systems could serve to disperse significant quantities of freshly synthesized LiBeB into the Interstellar Medium (ISM). It should be noted that there will be an additional contribution to Galactic LiBeB pollution from direct microquasar jet-ISM interactions and we report on this in a forthcoming paper (Butt et al., *in preparation*). Here we evaluate, with necessarily very approximate input parameters, the possibility that our proposed scenario may at least partially explain the anomalously high Li enhancements – roughly ~20-200 times the sun's photospheric value (Steenbock & Holweger 1984) – seen in the late-type companions of some BHXB systems (Martin et al. 1992, 1996). Such enhancements are very puzzling since Li nuclei are exceedingly fragile, being easily destroyed in the interiors of stars, and so Li would be expected to be depleted rather than enhanced in those stars. It is also worth mentioning that the genesis of LiBeB has not been resolved, and there is still no consensus on the dominant source of these light elements (eg. Ramaty et al., 1999).

## II. RELATIVE JET-ORBIT ORIENTATIONS

At present, observations of microquasar systems are not capable of constraining whether a given microquasar jet lies in the plane defined by the orbit of the binary companions. Measurements of the jet position angles (i.e. the angles in the plane of the sky) and jet inclination angles (i.e. the angles between the jets' directions and the lines of sight to the jets) can be deduced from radio measurements (Mirabel & Rodriguez 1999; Hjellming & Rupen 1995). This fixes their 3D geometry. In contrast, only the inclination angles of the binary planes are available from X-ray and optical orbital lightcurves. (The orbital inclination angle is defined as the angle between the line-of-sight and the normal to the orbital plane). If the position angle of a given orbital plane is roughly the same as the jet's position angle, then the jet could be directed sufficiently parallel to the binary plane as to periodically intercept the secondary star [Fig. 1]. The binary plane inclination angle required for eclipses to be seen is typically about 70º. Thus an offset between binary and jet inclination angles of more than about 70º degrees is a sufficient – *but not necessary* – condition for the jet to move closely enough to the binary plane to periodically intercept the secondary star. For example, the system V4641Sgr is constrained to have an angle of *at most* ~36º between its jet axis and binary plane just considering the relative inclination information ($i_{orbit}$=65º±5º ; Orosz et al., 2001 and $i_{jet}$≲6º; Orosz et al., 2001; Rupen, 2002) alone; the true (1$\sigma$) value lying between ~14º to ~36º, which certainly allows for the possibility of the jet striking the companion star.

Since the timescale for a black hole's spin to become aligned with the orbital angular momentum is typically longer than the lifetime of the binary system (Maccarone, 2002), and since a misalignment of a jet with respect to the binary plane of greater than about 70º will cause the jet to be intercepted by the secondary star, roughly about 10-20 % of jet systems are likely to have their jets periodically impacting the secondary stars, if the initial orientation angles are randomly distributed (assuming typical ratios of black hole to secondary mass of about 2-20). Indeed, theoretical considerations of supernova induced kicks indicate that large spin-orbit misalignments should be expected in NS microquasar systems (Brandt & Podsiadlowski, 1995). That the two microquasars where the jet and binary plane inclination angles are somewhat well established (GRO J1655-40 and V4641Sgr) have jets which are not perpendicular to their binary plane (Orosz et al., 2001; Rupen, 2002) suggests that the proposed initial misalignments cannot be too rare. Furthermore, it may be relevant that the accretion disks and jets of 'normal' extragalactic quasars (AGN) have recently been shown to clearly display similar misalignments (Schmitt et al. 2002).

Interestingly, Miller et al. (2002) have recently detected a broad Fe K$\alpha$ emission line in the *BeppoSAX* X-ray spectrum of V4641Sgr which they suggest implies an inclination of the inner accretion disk of $i_{\text{inner-disk}}$=43° ±15°. In our view, this measurement – if it can be confirmed with reduced observational and modeling uncertainties – may provide partial evidence of a hierarchy of inclinations which supports the picture of highly misaligned jet at the center of a warped accretion disk: $i_{\text{orbit}}$=65°±5° < $i_{\text{inner-disk}}$=43° ±15° < $i_{\text{jet}}$≲6°. (Although we note that systematic uncertainties in iron line measurement of 'inner' disk inclinations can be quite serious and that – within the error bars – the outer and inner disk inclinations could certainly be mutually consistent at a value of ~60° in V4641 Sgr). Hartnoll & Blackman (2000), have already shown how detailed spectroscopy of Fe K$\alpha$ emission lines could be used to probe details of disk warping, though their work was specific to AGN.

Assuming for simplicity that a jet is directed exactly parallel to the binary plane, one can compute the fraction of the time it will impact the secondary star. (In this preliminary study we consider only persistent jets as have been observed in several microquasar systems, although both persistent and transient behaviour may be displayed by the same system, eg. SS433). In the small angle approximation, the angle subtended by the secondary star will be its diameter divided by the orbital separation: for Roche lobe-filling systems this is given by ~ *0.92(1+q)$^{-1/3}$* radians, where *q=M$_x$/M$_2$* is the ratio of the mass of the compact star divided by the mass of the secondary star (Paczynski 1971). The fraction (*f$_{jet}$*) of the time the star is intercepting the jet is then simply twice this angle (since there is a jet and a counterjet) divided by the full *2π* radians of the orbit. We tabulate *f$_{jet}$*, the fraction of time one of the twin persistent jets is being intercepted, under the assumption that the jets lie in the orbital plane, for several system in table 1. A conservative typical value for this duty cycle is ~10 % of the time.

## III. Spallative Nucleosynthesis

In order to model the spallative yield on the surface of the secondary star one must know the material composition of both the jet and the surface of the companion, as well as the jet intensity (flux) and speed. Although optical data can tell us about the elemental composition of the secondary companion, the baryonic content of the jet remains largely unknown: models with even purely leptonic jets have been suggested (eg. Kaiser & Hannikainen 2002). Fortunately, in one microquasar system, SS433, high quality X-ray data has provided some insight into the nucleonic content of jets via the detection of red- and blue-shifted X-ray emission lines of Ne, Mg, Si, S, Ar, Ca, Fe and Ni which move in the same ephemeris as the jet precession (Kotani et al. 1994; Marshall, Canizares & Schulz, 2002), suggesting that baryonic jets are certainly not excluded by nature, and may even be the rule rather than the exception, given sufficient measurement sensitivity (eg. Heinz & Sunyaev 2002; see also Distefano et al. 2002 ).

We deduce a fiducial jet mass transfer rate from the total mass transfer rate given by Brown, Lee & Bethe (2001) of, $\dot{M}_j$~$5\times10^{-10}$ M$_\odot$ yr$^{-1}$ ~ $2\times10^{40}$ protons s$^{-1}$, which may be considered a conservative value in view of the jet mass transfer rate of SS433 of $\dot{M}_j$~$10^{-7}$ M$_\odot$ yr$^{-1}$ measured by Marshall, Canizares & Schulz (2002). Since, from the current observational standpoint, the persistent jets of SS433 are possibly the most highly baryonic and most powerful observed, our adopted rate is likely fairly representative for most microquasar systems and sufficiently accurate to at least make a rough assessment of our proposed scenario. (Although it is expected that the proposed jet-secondary impacts will in turn influence the mass transfer rate from the secondary, a detailed examination of this effect is beyond the scope of this preliminary study, and we thus assume a steady-state mass transfer rate for simplicity.) At typical BHXB orbital separations of 2-20 R$_\odot$, a jet with ~1$^\circ$ opening angle (eg. Marshall, Canizares & Schulz 2002) yields an irradiated surface of S~$10^{17} - 10^{20}$ cm$^2$ on the secondary. However, since we are interested in the steady-state *overall* surface LiBeB abundances, and not just the abundances of the exposed section of the secondary, we write the effective beam flux in terms of the jet illuminating the full cross-sectional area of the secondary star, and correcting for the jet impact duty cycle from table 1, we obtain:

$$F_{beam} \sim 8 \times 10^{16} \frac{\dot{m}_{-9}}{R_2 a_{10}} \; protons \; cm^{-2} \; sec^{-1} \qquad (1)$$

where $\dot{m}_{-9}$ is the jet mass ejection rate ($\dot{M}_j$) in units of $10^{-9}$ $M_\odot$/yr. $R_2$ is the companion radius expressed in $R_\odot$ and $a_{10}$ is the binary separation distance in units of $10 R_\odot$. Since most jets are observed to be substantially faster than SS433's ($\gtrsim 0.8c$ *vs.* 0.26c) we adopt a fiducial value of ~150MeV/n, corresponding to the average value of ~0.5c, as being representative and conservative: the faster the jet, the greater the amount of Li produced. Note that the relevant cross-sections are not very sensitive to the incident particle energy around ~150 MeV/n (MeV per nucleon) – the particle energy affects mainly the ionization range of the jet particles within the surface of the secondary and thus the 'skin-depth' of the secondary processed by the jet beam.

Given approximately solar-like initial abundances, the production of Li and other light isotopes occurs via two main channels [we consider only Li production here since its derived abundance can be directly compared with existing measurements, whereas Be and B spectroscopy is technically very difficult and thus those abundances are unavailable for comparison with any predictions (J. Orosz, *personal communication*)] :

$\alpha_{jet} + \alpha_{star} \rightarrow {}^{6,7}Li + X$

$p_{jet} + (CNO)_{star} \rightarrow {}^{6,7}Li + X$

the relevant cross-sections, $\sigma_{\alpha-\alpha}$ and $\sigma_{p\text{-}CNO}$, are ~1mbarn and ~10mbarn respectively in the $\gtrsim 100$ MeV/n energy range (1 barn=$10^{-24}$ cm$^2$). These two reactions contribute roughly equally to the production of Li because the product of the number densities $Y_\alpha$- $_{jet}$ × $Y_{\alpha\text{-}star}$ is ~10 times greater than the product $Y_{p\text{-}jet}$ × $Y_{CNO\text{-}star}$ for typical solar system abundances ([H/CNO]~1000 and [He/CNO]~100), which approximately compensates for the lower $\alpha$–$\alpha$ cross-section. We may then simply multiply the result for the p-CNO Li production rate by a factor of two to arrive at the total Li production rate.

The rate of $^{6,7}$Li production (per hydrogen atom of the secondary star within an ionization range of the beam particles) is then given by:

$$\frac{dY_{Li}}{dt} \sim 2\, F_{beam}\, \sigma_{p\text{-}CNO}\, Y_{CNO\text{-}star} \qquad (2)$$

Using the values above this may be evaluated to be:

$$\frac{dY_{Li}}{dt} \sim 10^{-9}\, Y_{CNO\text{-}star}\, \frac{\dot{m}_{-9}}{R_2 a_{10}} \quad \text{(H-atom sec)}^{-1} \qquad (3)$$

and the characteristic timescale of Li production is:

$$\tau_o \sim (F\sigma)_{total}^{-1} \sim 10^9\, \frac{R_2 a_{10}}{\dot{m}_{-9}} \quad \text{sec} \qquad (4)$$

On this timescale, roughly half the CNO and He nuclei exposed to the jet beam are transformed to Li. In general, the overabundance of these light nuclei (w.r.t. to the corresponding solar values) will depend on the duration $\Delta t$ of the irradiation:

$$\frac{Y_{Li}}{Y_{Li_\circ}} = \frac{Y_{CNO}}{Y_{Li_\circ}}\, \frac{\Delta t}{\tau_o} \qquad (5)$$

Considering an initially solar composition $(Y_{CNO}/Y_{Li})_\odot \sim 5 \times 10^5$, ones sees that even a relatively short continuous irradiation of only a few ~days may lead to a surface overabundances of Li of factors $\gtrsim 100$. However, the spallative production timescale must be compared to the relevant timescale for the *total* mass loss from the secondary star. In addition to the mass loss rate of the jet which is fed via accretion from the secondary, there is the (intrinsic and X-ray induced) wind driven mass-loss from the disk and secondary's envelope, as well as the mass loss rate to mass absorption by the compact object.

$$\dot{M}_{tot} = \dot{M}_j + \dot{M}_w + \dot{M}_c \qquad (6)$$

Observational (Boroson, Kallman & Vrtilek 2001; Vilhu 2002) and theoretical (Tavani & London 1993; Friend & Castor 1982) work in low- and high-mass XB systems indicates that the total mass loss rate, $\dot{M}_{tot}$, is dominated by the wind- and irradiation driven mass loss from the secondary and the disk and is roughly $\sim 10^{-7} - 10^{-6}$ M$_\odot$/yr, or in terms of scaled units: $(1-10) \times \dot{M}_{tot_{-7}}$.

Since the penetration depth of protons of energy E in the star is given by the ionization range, $X \sim E_{50}^2$ g cm$^{-2}$, where $E_{50}$ is the particle energy expressed in units of 50 MeV/n, the secondary's envelope mass processed via spallation is then:

$$M_{ENV} \sim 4\pi R_2^2 X \sim 3 \times 10^{-11} R_2^2 E_{50}^2 \quad M_\odot \qquad (7)$$

This mass is removed at a rate of, $\dot{M}_{tot}$, the total mass loss rate, on a timescale:

$$T \sim \frac{M_{ENV}}{\dot{M}_{tot}} \sim 10^4 R_2^2 E_{50}^2 \; \dot{M}_{tot_{-7}}^{-1} \quad \text{sec} \qquad (8)$$

Using $\Delta t = T$ in equation (5), we obtain:

$$\frac{Y_{Li}}{Y_{Li_\circ}} = \frac{Y_{CNO}}{Y_{Li_\circ}} \frac{10^4 R_2^2 E_{50}^2}{\tau_o} \; \dot{M}_{tot_{-7}}^{-1} \qquad (9)$$

Or, for typical values, $R_2 \sim 1$, $E_{50} \sim 3$, $a_{10} \sim 1$, $\dot{M}_{tot_{-7}} \sim 5$ & $\tau_0 \sim 10^9$ sec one should expect a steady state overabundance of Li (with respect to the average solar-system value) on the surface of the companion of a factor of ~8 (or ~$2\times10^{-5}$ $Y_{CNO}/Y_{Li_\odot}$), which is greater than but reasonably consistent with the observed values of $(Y_{Li}/Y_{Li_\odot}) \sim 0.1$-$1$ (Martin et al. 1992, 1996), especially considering that the observed values are, in fact, lower limits due to the ionizing effect of UV and X-ray irradiation (Martin et al. 1996).

It should be emphasized that the processed mass on the surface of the secondary may have been slightly overestimated above since the jets (and stars) carry magnetic fields and the mixing of the jet plasma and the stellar plasma may be suppressed to roughly one proton Lamor radius, which may be significantly less than the ionization 'skin-depth' we have used in our calculations for simplicity. Although the detailed treatment of such unknown magnetic geometries is clearly outside the scope of this study, we note that even if such an effect were to reduce the Li production efficiency by up to two orders of magnitude we would still retain a rough agreement with the observed abundances. A more important simplification in our analysis of jet-induced Li production has been to neglect some other effects which will accompany jet-secondary interactions. For example, the large power input to the secondary via the periodic impact of the jet may have a dramatic effect on its behaviour and evolution since most of the jet power will be turned to heat. This could lead to the secondary 'puffing-up' (eg. in a red-giant manner), and/or reradiating the energy. Indeed, it is possible that the recent enigmatic rapid optical and radio variability of V4641Sgr in radio outburst (Rupen, Dhawan & Mioduszewski, 2002) could have been the manifestation of an effect induced by the proposed jet-secondary interactions.

A possible observational signature of jet-induced Li production in certain microquasars may be a phase dependence in the strength of optical Li line, although this feature may be quite difficult to discern in practice.

It should be noted that we have (out of necessity) used only rough fiducial input parameters throughout for evaluating the feasibility of our proposed model. Future observational information on jet plasma compositions (ie. the nucleonic fraction) will certainly strongly constrain the nucleosynthetic aspect of our model, and we urge concerted observational efforts in this direction.

## IV. THE GAMMA-RAY LINES

Several gamma-ray features may be expected in the scenario described if they are produced in an optically thin region: nuclear line emission at 0.478 MeV from the deexcitation of $^7$Li* synthesized in the first excited state; the 2.2 MeV line from neutron capture of neutrons liberated via spallation; the 7.12, 6.92, 6.13, and 2.74 MeV nuclear lines from $^{16}$O*; the 4.44 MeV line of $^{12}$C*; and, at higher beam energies ($E_p$>280 MeV), the 67.5 MeV neutral pion-decay "hump" feature may also be present. However, with most of the instruments used to date (eg. SMM, TGRS, COMPTEL & OSSE) no evidence has been found for such lines except for a possible 2.22 MeV source in COMPTEL data (Harris et al., 1991,1997; McConnell et al. 1997; Harris et al., 2001) and a transient 481± 22 keV feature from BHXB Nova Muscae in ouburst by SIGMA/GRANAT (Goldwurm et al. 1992, Sunyaev et al. 1992). It is plausible that the latter gamma-ray line was, in fact, due to LiBeB nucleosynthesis in jet-secondary interactions, as described here. Tellingly, the companion star in Nova Muscae is known to have enhanced Li abundances (Martin et al. 1996). Observations of gamma-ray line signatures of the proposed process with INTEGRAL will be more sensitive and may prove more fruitful.

Since the companion stars are composed mostly of hydrogen and helium there is also expected to be significant neutron production via inelastic collisions. Some of these liberated neutrons will subsequently capture on the ambient nuclei, via, eg. n + $^1$H $\rightarrow$ $^2$H + $\gamma_{2.223MeV}$ - there can, however, also be non-radiative neutron capture on $^3$He to form $^4$He. If the neutrons are captured deep within the atmosphere of the secondary star then it is likely that the resulting 2.22MeV gamma-ray line will be Compton scattered before emerging from the atmosphere: the mean free path for a 2.22 MeV photon being ~10 g cm$^{-2}$. A detailed assessment of the neutron capture line intensity and shape, in the context of accretion generated neutrons, can be found in the very comprehensive study of Jean & Guessoum (2001).

If protons of energies greater than 280 MeV exist in the jet, inelastic collisions with the nuclei of the secondary star will produce one or more secondary mesons. The neutral pi-mesons, $\pi^o$'s, will promptly gamma-decay (98.8% of the time) to two photons of 67.5 MeV each in the pion rest frame. Due to the Lorentz transformation, in the observer's frame the spectrum is broadened and appears hump-like and symmetric about its peak at 67.5MeV (when plotted on a log $E\gamma$ scale; Stecker 1971). Such a feature may be detectable with the forthcoming GLAST satellite.

## VI. DISCUSSION & CONCLUSIONS

We have argued that a significant fraction of galactic microquasar systems' jets may be sufficiently misaligned with respect to their binary planes so as to impact their secondary stars. In this way these jets may play some role in synthesizing the light elements, Li, Be and B *in situ* on the secondary. We have shown that for reasonable – although necessarily very approximate – values of the parameter space we can reproduce the observed abundances of Li on the surfaces of several such systems. In this preliminary study we have assumed persistent jets throughtout and have neglected any extreme effects due to heating of the secondary star due to the jet power input. [That a microquasar jet may bend (as is seen in some extragalactic jets; eg. Conway & Murphy 1993, Conway & Wrobel, 1995) or that its position angle may be distorted due to relativistic aberration does not impact our hypothesis of possible jet-secondary impacts since the position angles of microquasar orbital planes remain ill-constrained.]

The fact that Li enhancements have been observed in the secondaries of virtually all XBs where they have been searched for – and for the stellar types where one expects to be able to discern the optical Li line – may either imply that jet-disk orientations are not random but that misalignment is the preferred state; or, it may argue for a separate process instead of, or in addition to, the proposed jet-secondary impacts as the source of the observed Li excess. For instance, even for well-aligned microquasars (as well as misaligned ones), it is plausible that surface irradiation of the secondary star by neutrons generated in the accretion disk may produce LiBeB via neutron spallation reactions as proposed by Guessoum & Kazanas (1999). However such a source is not thought to be a sufficiently prolific producer of Li as to be able to explain the high Li overabundances

observed – only in the month or so after a large outburst would such a mechanism be able to explain the observed Li abundances (Guessoum & Kazanas 1999). So such a steady-state process could then perhaps account for a certain minimum level of Li enrichment in XBs with the various possible geometries of jet-secondary impacts explaining the further enhancements and variations in those systems with especially high Li overabundances. Furthermore, as mentioned below, it is also possible that in both mis- and well-aligned microquasar systems that Li is generated in the local ISM via jet-ISM interactions (Butt et al., *in preparation*). If sufficient quantities of Li are formed close enough to secondary to be captured by it, this would result in a further steady-state Li enrichment mechanism. In the final measure however, the statistics of small numbers (six out of fourteen systems listed in Table I have known Li enhancements), together with the unknown distribution of jet-orbital orientations (ie. random *vs.* misalignment favored?), does not yet allow one to draw any firm conclusions about the nature of Li enrichment mechanism.

Mass loss from the accretion disk and envelope of the secondary star will serve to disperse the freshly generated nuclei into the ISM. Considering that the total Galactic population of LMXB and HMXB systems thought to be actively accreting at any one time is in the range $10^3$-$10^4$ (eg. Romani 1998), and that we expect ~10% of these to be sufficiently misaligned so as to permit jet-secondary collisions, we expect ~$10^{-13}$ $M_\odot$/yr of Li to be dispersed into the ISM via this process. However, analogously to what has been proposed already for AGN's by Famiano et al. (2001), these light nuclei will also be synthesized *in situ* in the ISM when microquasar jets impact the ambient media – we discuss this important aspect in a forthcoming paper (Butt et al., *in preparation*). Evidence of such jet-ISM/cloud interactions has, in fact, been already observed in the decelerating jets of XTE J1550-564 (Corbel 2002) and in SS433 (eg. Zealey, Dopita, & Malin, 1980; Fuchs 2002).

Interestingly, the envelopes of companion stars in Cataclysmic Variable (CV) binary systems (where a white dwarf accretes rather than a BH or NS) do not show enhanced Li abundances (Martin et al 1995): this implies that there is indeed something unique about the NS or BH systems which results in the enhanced Li abundances observed. We suggest that it may be the jets, which are seen only in the latter, microquasar, systems – though the hotter accretion disks of such systems could also be partly responsible via the mechanism proposed by Guessoum & Kazanas, 1999.

Our proposal here is premised on theoretical work (eg. Maccarone 2002; Brandt & Podsiadlowski 1995) which suggests that the jets and orbital planes of microquasars can be far from orthogonal, and for which there is good observational evidence already (Orosz et al., 2001; Rupen 2002; Miller et al., 2002). That normal extragalactic quasars, or AGN, clearly display such jet-disk misalignments (Schmitt et al 2002) may be interpreted as support of our hypothesis.

Direct measurements of the position angles of microquasar binary planes which will allow us to quantitatively assess the nature and statistics of misalignment, however, must await future missions such as the Space Interferometry Mission (Shao 1998), whose $10^{-5}$ arcsecond angular resolution should be sufficient to separate the primary and secondary stars. MAXIM, as proposed by Cash, White & Joy (2001), will have a $10^{-6}$ arcsecond angular resolution and ought to be able to separate the hot spot where the accretion stream hits the accretion disk and the inner disk itself as separate sources in the brightest X-ray binaries.

| Source | $M_x$ | $M_2$ | $q$ | $R_2/a$ | $f_{jet}$ |
|---|---|---|---|---|---|
| V404 Cyg | 12±2 | 0.6 | 20 | 0.167 | 0.11 |
| G2000+25 | 10±4 | 0.5 | 20 | 0.167 | 0.11 |
| N Oph 77 | 6±2 | 0.3 | 20 | 0.167 | 0.11 |
| N Mus 91 | 6±2.5 | 0.8 | 7.5 | 0.225 | 0.14 |
| A0620-00 | 10±5 | 0.6 | 17 | 0.176 | 0.11 |
| J0422+32 | 10±5 | 0.3 | 33 | 0.142 | 0.09 |
| J1655-40 | 6.9±1 | 2.1 | 3.3 | 0.284 | 0.18 |
| 4U1543-47 | 5.0±2.5 | 2.5 | 2 | 0.319 | 0.20 |
| Cen X-4 | 1.3±0.6 | 0.4 | 3 | 0.289 | 0.18 |
| 1915+105 | 14±4 | 1.2 | 12 | 0.198 | 0.13 |
| V4641Sgr | 10.4±1.7 | 6 | 1.5 | 0.340 | 0.22 |
| QZ Vul | ? | | | 0.159 | 0.10 |
| 1550-564 | 6.9±0.7 | 1.1 | 6.6 | 0.234 | 0.15 |
| 1118+480 | 6.9 | 0.5 | 14 | 0.2 | 0.12 |

Table 1: Using orbital information compiled in the review article by Charles (1998), we tabulate the fraction of the time each microquasar jet impacts its companion star ($f_{jet}$), under the extreme assumption that the jet lies in the plane of orbit. *a* is the binary separation distance and $q=M_x/M_2$ is the ratio of the mass of the compact star divided by the mass of the secondary star. If $R_2$ is defined as the equivalent radius of the companion's Roche lobe, then the angle subtended by the secondary star is given by 2 $R_2/a$ ~0.92 $(1+q)^{-1/3}$ radians (Paczynski 1971).

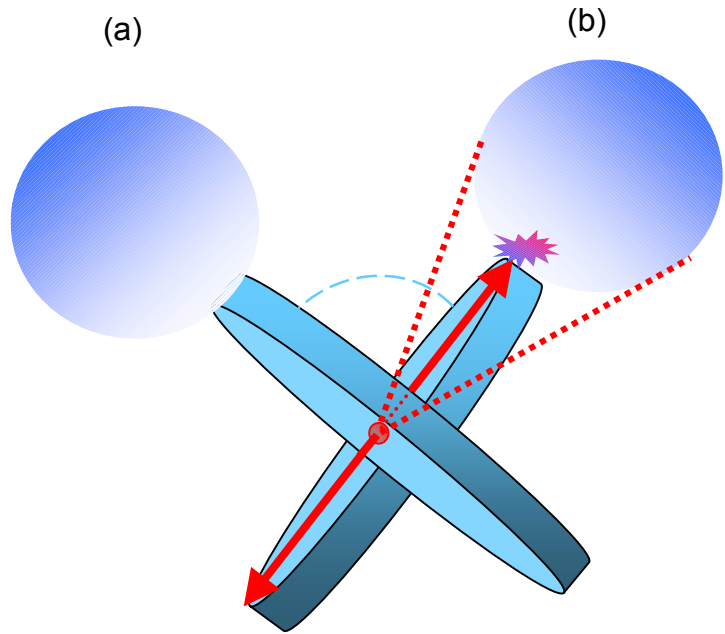

(a)  (b)

Fig 1: The position angle as well as the inclination angle of a given microquasar jet (thick red arrows) can be determined, which fixes its 3D geometry. In contrast, only the inclination angles of binary planes (disks in the schematic above) have thusfar been measured which introduces a degeneracy in the binary plane's possible position angles. Two possible <u>extreme</u> cases are shown above: (a) the "normal" orthogonal assumption; (b) the extreme case of the jet lying in the plane of orbit (shown only for illustration here: it being physically implausible that the jet lies *precisely* in the binary plane without disrupting the accretion disk). In case (b), within the range of relative position angles illustrated by the dotted red lines, the jet will strike the secondary star. Since the jet velocity is typically relativistic, it will initiate nuclear reactions on the secondary star, altering it's chemical content, and producing gamma-rays

We acknowledge useful conversations with J. McClintock, M. Harris, J. Orosz, P. Parker, M. Ribo and S. Safi-Harb. The comments of an anonymous referee are appreciated and improved the paper substantially. YMB acknowledges the support of the *CHANDRA* project, NASA Contract NAS8-39073.

Corresponding author Y.M.B. (e-mail: ybutt@cfa.harvard.edu)